# Ternary borides Nb$_7$Fe$_3$B$_8$ and Ta$_7$Fe$_3$B$_8$ with Kagome-type iron framework

Qiang Zheng,[*,a] Roman Gumeniuk,[a,b] Horst Borrmann,[a] Walter Schnelle,[a] Alexander A. Tsirlin,[d] Helge Rosner,[a] Ulrich Burkhardt,[a] Michael Reissner,[c] Yuri Grin,[a] and Andreas Leithe-Jasper[a]

Two new ternary borides $TM_7$Fe$_3$B$_8$ ($TM$ = Nb, Ta) were synthesized by high-temperature thermal treatment of samples obtained by arc-melting. This new type of structure with space group $P6/mmm$, comprises $TM$ slabs containing isolated planar hexagonal [B$_6$] rings and iron centered $TM$ columns in a Kagome type of arrangement. Chemical bonding analysis in Nb$_7$Fe$_3$B$_8$ by means of the electron localizability approach reveals two-center interactions forming the Kagome net of Fe and embedded B, while weaker multicenter bonding present between this net and Nb atoms. Magnetic susceptibility measurements reveal antiferromagnetic order below $T_N$ = 240 K for Nb$_7$Fe$_3$B$_8$ and $T_N$ =265 K for Ta$_7$Fe$_3$B$_8$. Small remnant magnetization below 0.01 $\mu_B$/f.u. is observed in the antiferromagnetic state. The bulk nature of the magnetic transitions was confirmed by the hyperfine splitting of the Mößbauer spectra, the sizable anomalies in the specific heat capacity, and the kinks in the resistivity curves. The high-field paramagnetic susceptibilities fitted by the Curie-Weiss law show effective paramagnetic moments $\mu_{eff} \approx$ 3.1 $\mu_B$/Fe in both compounds. The temperature dependence of the electrical resistivity also reveals metallic character of both compounds. Density functional calculations corroborate the metallic behaviour of both compounds and demonstrate the formation of a sizable local magnetic moment on the Fe-sites. They indicate the presence of both antiferro- and ferrromagnetic interactions.

## Introduction

Transition-metal borides are remarkable for their physical, chemical and mechanical properties, in particular, combining refractory behavior, high hardness, chemical inertness, and metallic conductivity.[1-7] To exemplify a few well-known materials of high technological relevance, there are boride-containing metallic glasses,[8] LaB$_6$ cathodes for electron microscopes,[9] and Nd$_2$Fe$_{14}$B-based permanent magnets.[10] On the other hand, soft magnetic properties particularly found in ferrous amorphous boron-containing alloys have led to several highly useful applications, such as electro-magnetic materials.[11, 12] The structural complexity of electron-deficient boron and its compounds is caused by the intricate ways how their valence requirements are satisfied.[13-15] Therefore, in metal-borides this not only gives rise to the formation of one-, two-, or three-dimensional arrangements of covalently bonded boron atoms,[16-21] but also, due to complex structures, results in a multitude of physical interactions which can lead to superconductivity[22-26] or magnetism.[27-31]

In a crystal structure, these interactions are governed by the spatial arrangement and coordination of the constituents which carry a magnetic moment. Accordingly, in borides phases with exotic and complex magnetic ground states can be expected and merit explorative research. To this purpose, we investigated the ternary $TM$–Fe–B ($TM$ = Nb, Ta) systems which have been studied since 1960s. The hitherto known ternary compounds in these two systems are $TM$FeB, $TM_2$FeB$_2$, $TM_3$Fe$_3$B$_4$ and TaFeB$_3$.[32-35] We attempted to synthesize $TM_3$Fe$_3$B$_4$ but failed. Instead, the analysis of samples with nominal compositions $TM_3$Fe$_3$B$_4$ revealed the appearance of new compounds $TM_7$Fe$_3$B$_8$, which crystallize in a primitive hexagonal lattice. This new type of structure comprises $TM$ slabs containing planar hexagonal boron rings and iron centered $TM$ columns in a Kagome type of arrangement (see below).

The interest in the Kagome lattices of magnetic ions is triggered by their strongly frustrated nature. No ordered antiferromagnetic magnetic configuration can be stabilized in such a geometry, and an exotic spin-liquid ground state is formed instead.[36, 37] However, in order to suppress any ordering a two-dimensional isotropic next-neighbor coupling is the prerequisite. Systems with the Kagome-like arrangement of magnetic ions range from Cu$^{2+}$ minerals[38, 39] to Ce-based intermetallic compounds,[40] but only few of them reveal the anticipated spin-liquid state at low temperatures.[41-43] Even subtle geometrical distortions or interactions beyond nearest neighbors are usually sufficient to alleviate the frustration and stabilize the magnetic order.[44, 45]

The majority of the kagome-lattice compounds reported so far are magnetic insulators. Kagome lattices in itinerant systems are by far more exotic, and no magnetic metals with the ideal kagome geometry have been reported to date. In the following, we fill this gap by investigating the $TM_7$Fe$_3$B$_8$ borides, where three-fold symmetry of the crystal structure ensures perfect frustration on individual triangles of the kagome lattice. However, these compounds are magnetically ordered with relatively high Néel temperatures. We suggest that strong interplane couplings arising from the geometrical proximity of the kagome planes may be instrumental in stabilizing the

[a.] Max-Planck-Institut für Chemische Physik fester Stoffe, Nöthnitzer Str. 40, 01187 Dresden, Germany.
[b.] Institut für Experimentelle Physik, TU Bergakademie Freiberg, Leipziger Str. 23, 09596 Freiberg, Germany.
[c.] Institut für Festkörperphysik, TU Wien, Wiedner Hauptstr. 8-10, 1040 Wien, Austria.
[d.] Experimentalphysik VI, EKM, Institut für Physik, Universität Augsburg, D-86135 Augsburg, Germany.
[*] Corresponding author, E-mail address: zheng@cpfs.mpg.de; qiangzhengsic@gmail.com. Present address: Materials Science & Technology Division, Oak Ridge National Laboratory, Oak Ridge, TN 37831, USA
Electronic Supplementary Information (ESI) available: [details of any supplementary information available should be included here]. See DOI: 10.1039/x0xx00000x

Table 1. Crystallographic Data for $Nb_7Fe_3B_8$ and $Ta_7Fe_3B_8$

| Composition | $Nb_7Fe_3B_8$ | $Ta_7Fe_3B_8$ |
|---|---|---|
| Crystal system | hexagonal | |
| Space group | $P6/mmm$ (No. 191) | |
| Lattice parameters | | |
| $a$ (Å) | 8.3346(2) | 8.2788(3) |
| $c$ (Å) | 3.2941(1) | 3.2934(3) |
| $V$ (Å$^3$) | 198.17(2) | 195.48(3) |
| Formula unit/cell, $Z$ | 1 | 1 |
| Calculated density/(g cm$^{-3}$) | 7.5777(6) | 12.916(2) |
| Crystal size (mm$^3$) | 0.02×0.02×0.02 | 0.04×0.04×0.04 |
| Diffraction system | Rigaku R-Axis Rapid | Rigaku AFC7 |
| Radiation; $\lambda$ (Å) | Mo $K_\alpha$; 0.71073 Å | |
| $2\theta_{max}$(°) | 131.69 | 66.40 |
| Absorption coefficient (mm$^{-1}$) | 14.751 | 108.146 |
| $N(hkl)_{measured}$ | 5932 | 1774 |
| $N(hkl)_{unique}$ | 719 | 179 |
| $N(hkl)_{observed}$ ($F_{hkl} > 4\sigma(F)$) | 614 | 174 |
| $R_{int}/R_\sigma$ | 0.050/0.023 | 0.045/0.016 |
| Refined parameters | 17 | 14 |
| $R_F/wR_F^2$ | 0.047/0.053 | 0.017/0.017 |
| Extinction coefficient | 0.063(4) | 0.026(1) |

magnetically ordered state in these novel ternary Fe-containing borides.

## Experimental section

### Sample preparation

The following elements were used to prepare the samples: Ta and Nb foil (Chempur, 99.9 mass%), Fe foil (Alfa Aesar, 99.999 mass%) and crystalline B powder (Alfa Aesar, 99.999 mass%). The mixtures of the elements with compositions $TM_3Fe_3B_4$ and $TM_7Fe_3B_8$ were arc-melted several times to obtain homogeneous samples, and the mass losses during arc-melting were 1–3.3 %. The obtained ingots were placed inside $Al_2O_3$ crucibles and then sealed in Ta tubes. The annealing was carried out at 1000 °C for 12h and then 1500 °C for 48h in a high-temperature furnace (HTM Reetz, LORA). Additionally, all above manipulations were performed inside argon-filled glove boxes ($p(O_2/H_2O) \leq 1$ ppm). The resulting samples are stable in air for a long time.

### Powder and single-crystal X-ray diffraction

Powder X-ray diffraction (XRD) data were collected on a HUBER G670 imaging plate Guinier camera equipped with Co $K_{\alpha 1}$ radiation ($\lambda$ = 1.78897 Å). Phase analysis and indexing were performed using the *WinXPow* program package.[46] Lattice parameters were refined by least-squares fitting with $LaB_6$ internal standard within the program package *WinCSD*.[47]

$TM_7Fe_3B_8$ single crystals were selected from the samples with the nominal compositions $TM_3Fe_3B_4$. Single crystal diffraction data were collected on a R-Axis Rapid or Rigaku AFC7 diffractometer equipped with Mercury CCD detectors (Mo $K_\alpha$ radiation, $\lambda$ = 0.71073 Å). Absorption correction was made using a multi-scan procedure. The crystal structures were solved by a direct phase determination method and refined by a full-matrix least-squares procedure within the program package *WinCSD*.[47] Details on the single-crystal diffraction data collection and structural refinement are listed in Table 1.

### Metallography

Pieces with several millimeters size were cut from the annealed samples for metallographic investigations. They were embedded in conductive resin and then subjected to a multistep grinding and polishing process to achieve high-quality polished surfaces. The microstructures were investigated by optical microscopy (Axioplan2, Zeiss) as well as scanning electron microscopy (Philips XL 30 with a $LaB_6$ cathode, FEI). The chemical compositions were analyzed by means of energy dispersive X-ray spectroscopy (EDXS, Philips XL 30) and wavelength dispersive X-ray spectroscopy (WDXS, Cameca SX 100, W cathode, S-UTW-Si-(Li) detector).

The determination of boron content by microprobe WDXS is challenging due to the general issue that the measured intensities are related to the mass concentrations, while boron is extremely light as compared to niobium, tantalum, and iron. Also, the very low energies of boron X-ray lines give rise to a strong influence of absorption effects. Therefore, completely detected intensities originating from the uppermost surface layer and its extending area are strongly influenced by the quality of the local area that is probed by the electron beam, and the energy and shape of the boron X-ray lines are influenced significantly by the local chemical environment and bonding situation of boron. For this reason, reference compounds should be materials, in which boron should have the similar chemical environment and bonding situation with the two analyzed compounds. Subsequently, single crystals of NbFeB, TaFeB in sizes of several millimeters were successfully grown in melt fluxes and satisfy requirement for WDXS reference materials. By applying ICP-MS technique, their analyzed compositions are $Nb_{1.00(1)}Fe_{1.02(1)}B_{1.01(2)}$ and $Ta_{1.00(1)}Fe_{1.03(1)}B_{1.01(1)}$, respectively, revealing both are stoichiometric.[48] The intensities of Ta $L_\alpha$, Fe $K_\alpha$ and B $K_\alpha$ lines were measured for the Ta-containing compound, while due to the overlapping of B $K_\alpha$ line with Nb $M_\gamma$ line, the intensities of Nb $L_\alpha$, Fe $K_\alpha$ and B $K_{\alpha 2}$ lines were measured for the Nb-containing compound.

The matrix correction model according to Pouchou and Pichoir[49] was applied to calculate the chemical compositions. Different conditions have been applied for the measurement of the x-ray lines of the heavy elements and boron. For the Ta-containing compound, currents of 15 nA and 40 nA under an acceleration voltage of 20 kV were applied for the measurement of the intensities of Ta $L_\alpha$ and Fe $K_\alpha$ lines, respectively, with the dwelling time of 3 seconds for each position. The intensity of boron $K_\alpha$ line was measured by the area intensity method, with larger current of 100 nA, acceleration voltage of 7 kV and dwelling time 3 seconds for each position. For the Nb compound, the intensities of Nb $L_\alpha$ and Fe $K_\alpha$ lines was measured by applying a current of 60 nA under an acceleration voltage of 15 kV with dwelling time 3 seconds for each position. The intensity of B $K_{\alpha 2}$ line was measured by applying the same acceleration voltage and current used for the Ta compound, however, due to the much weaker intensity of B $K_{\alpha 2}$ line, dwelling time for each position was 1136 seconds.

**Transmission electron microscopy (TEM) observations**

Electron diffraction and high-resolution TEM (HRTEM) observations were both performed using a field-emission electron microscope JEM 2100F (JEOL, Japan) operating at 200 kV. HRTEM image simulations were carried out with program STEM_CELL.[50]

**Physical Properties**

Magnetization at external magnetic fields $\mu_0H$ ranging from 0.01 T to 7 T (temperature range 1.8 K–400 K) was measured in a SQUID magnetometer (MPMS XL-7, Quantum Design) on polycrystalline samples. The electrical resistance was recorded by a four contact method using low-frequency alternating current (ACT option, PPMS, Quantum Design) on small bar-shapes pieces in zero field and in a field $\mu_0H$ = 9 T. Heat capacity was determined by a relaxation method (HC option, PPMS, Quantum Design) in fields $\mu_0H$ of 0, 3, 6, and 9 T.

**Mossbauer spectroscopy**

$^{57}$Fe Mössbauer spectra were recorded at 294 K and 4.3 K. The measurements were performed with a standard constant acceleration spectrometer in transmission geometry in a continuous flow cryostat with the sample kept at helium atmosphere. The $^{57}$CoRh source was mounted on the driving system and kept at room temperature. All center shift (CS) data are given relative to this source. Calibration of the velocity scale was carried out with $\alpha$-Fe foils. The spectra were analysed by solving the full Hamiltonian including electrostatic and magnetic hyperfine interactions. Sample thickness was taken into account by the method of Mørup and Both.[51]

**Electronic structure calculations**

The electronic structure of $TM_7Fe_3B_8$ was calculated within the framework of density functional theory (DFT) using the full-potential code FPLO.[52] The local density approximation (LDA) to the exchange-correlation potential was chosen.[53] Reciprocal space was sampled by a fine $k$-mesh with 630 points in the symmetry-irreducible part of the first Brillouin zone for the crystallographic unit cell of $TM_7Fe_3B_8$ and 190 points for the supercell doubled along the $c$ direction. Convergence with respect to the $k$-mesh was carefully checked. For the spin-polarized calculations, the highest crystallographic symmetry compatible with the magnetic ordering pattern was used in order to facilitate the convergence.

**Chemical bonding analysis**

Analysis of chemical bonding was performed for $Nb_7Fe_3B_8$ using the lattice parameters and atomic coordinates from the crystal structure refinement of single-crystal X-ray diffraction data (Tables 1 and 2). The TB-LMTO-ASA program package[54] was employed using the Barth-Hedin exchange potential[55] for the LDA calculations. The radial scalar-relativistic Dirac equation was solved to obtain the partial waves.[56] Addition of empty spheres was not necessary because the calculation within the atomic sphere approximation (ASA) includes corrections for the neglect of interstitial regions and partial waves of higher order.[57] The following radii of the atomic spheres were applied for the calculations $r$(Nb1) = 1.69 Å, $r$(Nb2) = 1.49 Å, $r$(Fe) = 1.48 Å, $r$(B1) = 1.02 Å, $r$(B2) = 1.19 Å. A basis set containing Nb(5$s$,5$p$,4$d$), Fe(4$s$,4$p$,3$d$) and B(2$s$,2$p$) orbitals was employed with Nb(4$f$) and B(3$d$) functions being downfolded.

The electron localizability indicator (ELI, $\Upsilon$) was evaluated in the ELI-D representation[58-60] with an ELI-D module within the TB–LMTO–ASA program package. Topological analysis of the electron localizability indicator, e.g., localization of the ELI maxima as indicators of the direct atomic interactions, estimation of their basins were performed with the program DGrid.[61]

## Results and discussion

**Phase formation and crystal structure determination**

Both as-cast samples with the nominal compositions $TM_3Fe_3B_4$ and $TM_7Fe_3B_8$ contain $Fe_2B$, $TM$B and a little amount of other $TM$–B binary phases, however, with no traces of any other ternary phases. The $TM_7Fe_3B_8$ compounds were firstly discovered as matrix phases in the samples with nominal composition $TM_3Fe_3B_4$ annealed at 1500 °C for 48 h. Subsequently, bulk samples of $TM_7Fe_3B_8$ with minor amounts of impurity phases were obtained by annealing at 1500 °C for 48h. However, both compounds decomposed into $TM$B, $Fe_2B$ phases (and sometimes with α-Fe) at temperatures of about 1600 °C. The decomposition was proved by an endothermal peak in DSC curves at ~1580 °C for the $Nb_7Fe_3B_8$ sample.

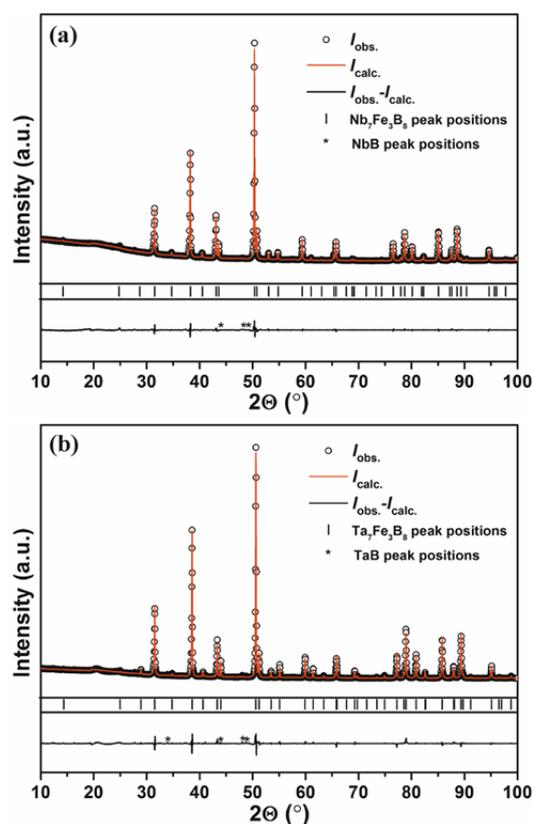

**Figure 1.** Powder XRD patterns of the samples with nominal composition (a) $Nb_7Fe_3B_8$ and (b) $Ta_7Fe_3B_8$ annealed at 1500 °C.

Powder XRD patterns of the sample $TM_7Fe_3B_8$ annealed at 1500 °C are shown in Fig. 1. 26 strongest reflections in the Nb-containing sample were indexed successfully in a hexagonal primitive lattice with unit cell parameters $a$ = 8.3346(2) Å, $c$ = 3.2941(1) Å (Figure of merit (FOM) = 72.7) using the automatic indexing algorithm TREOR within WinXPow program package. In the same way, 23 strongest reflections in the Ta-containing sample were also successfully indexed in a hexagonal primitive unit cell with lattice parameters $a$ = 8.2788(3) Å, $c$ = 3.2934(3) Å (FOM = 78.2). No extinction conditions were observed in the two data sets. The remaining reflections belong to the $TM$B phases (~1 vol. % and ~3 vol. % in $Nb_7Fe_3B_8$ and $Ta_7Fe_3B_8$, respectively).

The results of the single-crystal X-ray diffraction experiment are well consistent with the powder pattern indexing. Analysis of the reflection conditions revealed Laue class 6/$mmm$, and the most symmetric space group $P6/mmm$ was chosen to solve the crystal structures. Since light B atoms give only a small contribution to the reflection intensities, reflexes were measured up to high angles ($2\theta_{max}$ = 131.69°) for the Nb-containing single crystal in order to facilitate the determination of the positions of B atoms. The initial positions of Nb/Ta and Fe atoms were obtained by the direct phase determination method, while the positions of B atoms were found from difference Fourier maps. The refinement was carried out with anisotropic approximation for the atomic displacement parameters for all the atoms (except B atoms in $Ta_7Fe_3B_8$ where refinement due to relatively large absorption is questionable). Occupations of the Ta, Nb and Fe positions were also refined, revealing that all are fully occupied. Crystallographic data and final atomic coordinates as well as isotropic displacement parameters are listed in Table 1 and Table 2. Anisotropic displacement parameters are deposited in Table S1 in the Supplementary Information. Compositions are in good agreement with WDXS results. By WDXS, the compositions for the two compounds were achieved as follows: $Ta_{7.00(5)}Fe_{3.13(3)}B_{7.92(7)}$ and $Nb_{7.00(6)}Fe_{3.10(3)}B_{7.4(2)}$. They are satisfactorily close to the compositions $TM_7Fe_3B_8$ obtained from single crystal diffraction data, and the former one is nearly the same. The lower precision and accuracy of the obtained boron content for the Nb-containing compound

**Table 2.** Atomic coordinates and isotropic displacement parameters for $Nb_7Fe_3B_8$ and $Ta_7Fe_3B_8$

| atom | site | x | y | z | $B_{iso}/B_{eq}$[a] |
|---|---|---|---|---|---|
| $Nb_7Fe_3B_8$ | | | | | |
| Nb1 | 1$a$ | 0 | 0 | 0 | 0.30(2) |
| Nb2 | 6$l$ | 0.21202(3) | 2$x$ | 0 | 0.313(8) |
| Fe | 3$g$ | 1/2 | 0 | 1/2 | 0.41(2) |
| B1 | 2$d$ | 1/3 | 2/3 | 1/2 | 0.5(1) |
| B2 | 6$k$ | 0.2186(8) | 0 | 1/2 | 0.51(9) |
| $Ta_7Fe_3B_8$ | | | | | |
| Ta1 | 1$a$ | 0 | 0 | 0 | 0.20(2) |
| Ta2 | 6$l$ | 0.21233(3) | 2$x$ | 0 | 0.27(2) |
| Fe | 3$g$ | 1/2 | 0 | 1/2 | 0.34(5) |
| B1 | 2$d$ | 1/3 | 2/3 | 1/2 | 0.3(1) |
| B2 | 6$k$ | 0.221(1) | 0 | 1/2 | 0.4(1) |

[a] $B_{eq} = 1/3[a^{*2}a^2B_{11} + b^{*2}b^2B_{22} + c^{*2}c^2B_{33} + 2aba^*b^*(\cos\gamma)B_{12} + 2aca^*c^*(\cos\beta)B_{13} + 2bcb^*c^*(\cos\alpha)B_{23}]$

results mainly from using the much weaker B $K_{\alpha2}$ line. Moreover, WDXS compositions are consistent with the results by standardless EDXS analysis, which yield the Ta:Fe and Nb:Fe atomic ratios 70.6(8):29.4(8) and 70(1):30(1), respectively.

**TEM investigation**

The TEM study of $Ta_7Fe_3B_8$ confirmed the results of the crystal structure determined by the X-ray diffraction. The electron diffraction patterns along relevant zone axes are shown in Fig. 2. All five patterns can be well indexed in a hexagonal primitive lattice with the cell parameters obtained from powder XRD diffraction (Table 1). There are no superstructure reflections or diffuse reflections characteristic for disorder.

The atomic arrangement in the $TM_7Fe_3B_8$ structure determined from single crystal diffraction (as shown in Fig. 4a) is proved by HRTEM observations (Fig. 3). For the image along [100] (Fig. 3a), the simulated image (the inset in Fig. 3a) was calculated at defocus $\Delta f$ = −40 nm and thickness $t$ = 6.7 nm, in which the dark contrast mainly comes from the projection of Ta and Fe atoms. For the HRTEM image along [001] (Fig.3b; the inset shows the simulated image at a defocus $\Delta f$ = −20 nm and thickness $t$ = 6.6 nm), the bright spots are caused by the projection of Ta1 and Fe atoms. HRTEM images also reveal no extended defects in $Ta_7Fe_3B_8$.

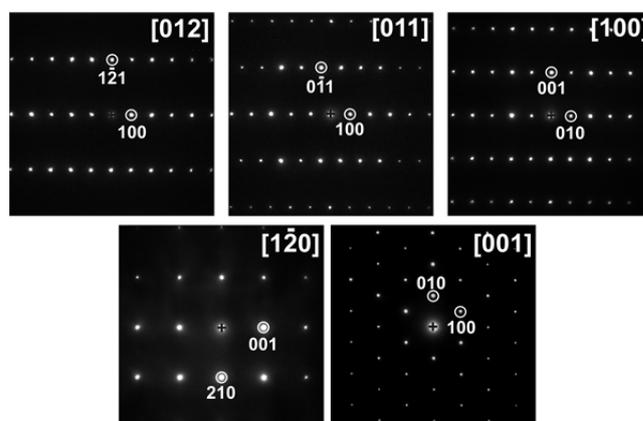

**Figure 2.** Electron diffraction patterns for $Ta_7Fe_3B_8$ along different zone axes.

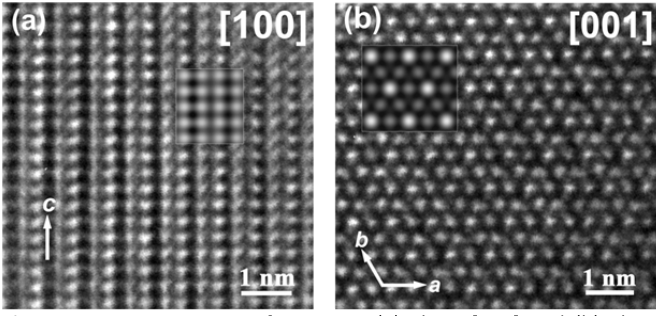

**Figure 3.** HRTEM images of Ta$_7$Fe$_3$B$_8$: (a) along [100] and (b) along [001] (the two insets within white lines are the simulated images for $\Delta f = -40$ nm and $t = 6.7$ nm along [100] and $\Delta f = -20$ nm and $t = 6.6$ nm along [001], respectively).

**Crystal structure**

The crystal structures of $TM_7$Fe$_3$B$_8$ adopt a new type of atomic arrangement. As shown in Fig. 4, the structures consist of two alternating layers: the layer of the $TM$ atoms at $z = 0$ and the layer of the B and Fe atoms at $z = 1/2$. The B and Fe atoms reside inside [$TM_6$] trigonal prisms and center [($TM2)_8$] tetragonal prisms, respectively. The trigonal prisms comprise two different types: one is represented by [(B2)($TM1)_2$($TM2)_4$], and the other one by [(B1)($TM2)_6$], respectively. Both represent typical structural building blocks observed in intermetallic borides. The former type of the polyhedra is corrugated and forms hexagonal columns along [001]. Such kind of hexagonal columns has been previously observed in the ternary borides $TM_7TM'_6$B$_8$ ($M$ = Nb, Ta; $M'$ = Ru, Rh, Ir)[62] and their disordered variant Ti$_7$Rh$_4$Ir$_2$B$_8$.[63] In $TM_7$Fe$_3$B$_8$, the hexagonal columns are separated by Fe and B1 atoms. This gives rise to the formation of [Fe($TM2)_8$] tetragonal prisms and the second type of trigonal prisms. Such buildup of trigonal prims and tetragonal prims in $TM_7$Fe$_3$B$_8$ structures is observed in borides for the first time. This type of structure pattern was previously found in the crystal structure of BaFe$_2$Al$_9$.[64] Here the aluminum atoms are occupying the positions of $TM2$ and Fe, iron atoms are located at the B2 positions, barium is shifted by [00$^1/_2$] in respect to Nb1 site, and the positions of B2 are not occupied: Ba$_1$Al$_6$Al$_3$Fe$_2\square_6$ is equivalent to $(TM1)_1(TM2)_6$Fe$_3$(B1)$_2$(B2)$_6$. Later another ordering variant for this atomic motif was discovered in Hf$_5$Nb$_5$Ni$_3$P$_5$: Hf$_1$(Hf,Nb)$_6$(Hf,Nb)$_3$P$_2$(Ni,P)$_6$.[65]

**Table 3.** Selected interatomic distances (Å) in the structures of Nb$_7$Fe$_3$B$_8$ and Ta$_7$Fe$_3$B$_8$; CN= coordination number.

| Atoms | | Nb$_7$Fe$_3$B$_8$ | Ta$_7$Fe$_3$B$_8$ | CN |
|---|---|---|---|---|
| $TM1$ | $-12$B2 | 2.456(3) | 2.462(3) | 20 |
| | $-6TM2$ | 3.0607(4) | 3.0447(2) | |
| | $-2TM1$ | 3.2941(1) | 3.2934(3) | |
| $TM2$ | $-4$B2 | 2.396(2) | 2.383(6) | 17 |
| | $-2$B1 | 2.4041(3) | 2.3921(3) | |
| | $-4$Fe | 2.7120(2) | 2.6995(3) | |
| | $-2TM2$ | 3.0333(6) | 3.0053(6) | |
| | $-1TM1$ | 3.0607(4) | 3.0447(2) | |
| | $-2TM2$ | 3.0607(4) | 3.0447(2) | |
| | $-2TM2$ | 3.2941(1) | 3.2934(3) | |
| Fe | $-2$B2 | 2.345(3) | 2.310(4) | 14 |
| | $-2$B1 | 2.4060(1) | 2.3899(1) | |
| | $-8TM2$ | 2.7120(2) | 2.6995(3) | |
| | $-2$Fe | 3.2941(1) | 3.2934(3) | |
| B1 | $-6TM2$ | 2.4041(3) | 2.3921(3) | 9 |
| | $-3$Fe | 2.4060(1) | 2.3899(1) | |
| B2 | $-2$B2 | 1.822(5) | 1.830(6) | 9 |
| | $-1$Fe | 2.345(3) | 2.310(4) | |
| | $-4TM2$ | 2.396 (2) | 2.383(6) | |
| | $-2TM1$ | 2.456(3) | 2.462(3) | |

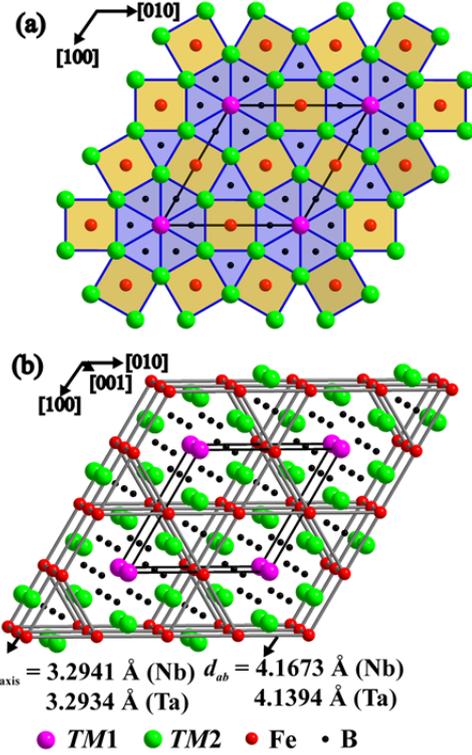

**Figure 4.** Crystal structures of $TM_7$Fe$_3$B$_8$: (a) framework and packings of polyhedral prisms (pink – $TM1$ at 1$a$ site, green – $TM2$ at 6$l$ site, red – Fe at 3$g$ site, black – B1 at 2$d$ site and B2 at 6$k$ site; blue – [B$TM_6$] trigonal prism, yellow – [Fe$TM_8$] tetragonal prism); (b) three-dimensional Fe–Fe framework and Fe–Fe distances in the (001) plane and along [001].

The structural motif of $TM_7$Fe$_3$B$_8$ can be described as a 2D intergrowth of the AlB$_2$-type ($RX_2$ fragment) and CsCl-type ($RR'$ fragment) slabs with the general formula $R_{m+n}R'_mX_{2n}$[66] where $m$ = 3 and $n$ = 4 are the numbers of outlined structural fragments per unit cell of $TM_7$Fe$_3$B$_8$. In the AlB$_2$-type $TM$B$_2$ structures,[67, 68] condensed [B$TM_6$] trigonal prisms form 3D blocks completely filling the space. Due to such an arrangement, the boron atoms come in close contact and form infinite two dimensional graphite-like nets. Interestingly, in the $TM_7$Fe$_3$B$_8$ structure, the planar six-membered rings of boron are now separated by CsCl-type $TM$Fe slabs, and this also additionally results in the isolated B1 atoms in this structure. Such kind of isolated planar [B$_6$] rings and the isolated B atoms were also observed in the ternary borides $TM_7TM'_6$B$_8$[62] and their disorder variant Ti$_7$Rh$_4$Ir$_2$B$_8$.[63] Moreover, the $TM_7$Fe$_3$B$_8$ structure is also related to the Mo$_2$FeB$_2$-type $TM_2$FeB$_2$ structure,[33, 34] which can also be described as an intergrowth of the AlB$_2$-type and CsCl-type slabs, now with $m$ = 2 and $n$ = 2 in $R_{m+n}R'_mX_{2n}$. One nearly single-phase Ta$_2$FeB$_2$ sample was also synthesized during this study and its structure was refined from powder XRD data.[69] In $TM_2FeB_2$, the boron atoms centering the trigonal prism sites come together in $B_2$-pairs, while in $TM_7Fe_3B_8$, these pairs are fused into [$B_6$] rings or broken up into isolated B atoms.

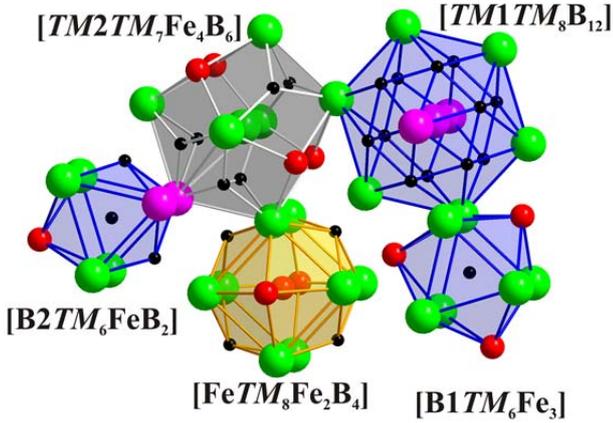

**Figure 5.** Coordination polyhedra of the *TM*, Fe and B atoms in $TM_7Fe_3B_8$.

Coordination polyhedra of all the atoms in $TM_7Fe_3B_8$ are depicted in Fig. 5. The polyhedron of *TM*1 is a 20-vertices polyhedron consisting of 12 B and 8 *TM* atoms, which could also be observed for *TM* atoms in $TM_7TM'_6B_8$,[62] $TMB_2$,[67,68] and $TM_3B_4$.[70] *TM*2 and Fe are coordinated by a 17-vertices [$TM_7Fe_4B_6$] polyhedron and a six-capped [$TM_8Fe_2B_4$] cube, respectively. The same coordination polyhedra for *TM* as well as Fe are found in $TM_2FeB_2$.[33, 34] Finally, B atoms are coordinated by tricapped trigonal prisms formed either by 6 *TM* and 3 Fe atoms or by 6 *TM*, 1 Fe and 2 B atoms.

Selected interatomic distances in the crystal structures of $TM_7Fe_3B_8$ compounds are listed in Table 3. All of them are close or slightly larger than the sum of atomic radii of the elements ($r_{Ta}$ = 1.43 Å, $r_{Nb}$ = 1.43 Å, $r_{Fe}$ = 1.24 Å, $r_B$ = 0.83 Å).[71] *TM*–*TM* contacts in the (001) plane for $Nb_7Fe_3B_8$ (3.0333(6) Å and 3.0607(4) Å) and $Ta_7Fe_3B_8$ (3.0053(6) Å and 3.0447(2) Å) are both shorter than these distances in the (001) plane in $TM_7Ru_6B_8$ (3.1257 Å and 3.1189 Å for $Nb_7Ru_6B_8$ and $Ta_7Ru_6B_8$, respectively)[62] and $AlB_2$-type $TMB_2$ (3.1115 Å and 3.076 Å for $NbB_2$,[67] and $TaB_2$,[68] respectively). *TM*–*TM* contacts along the [001] direction for $Nb_7Fe_3B_8$ (3.2941(1) Å) and $Ta_7Fe_3B_8$ (3.2934(3) Å) are both longer than these distances along [001] in $TM_7Ru_6B_8$ (3.1284(3) Å and 3.1370(3) Å for $Nb_7Ru_6B_8$ and $Ta_7Ru_6B_8$, respectively)[62] and $TMB_2$ (3.2657 Å and 3.275 Å for $NbB_2$,[67] and $TaB_2$,[68] respectively). B2 atoms in the isolated hexagonal [$B_6$] rings in $TM_7Fe_3B_8$ reside slightly off-center in the trigonal prisms towards the Fe atom caps, resulting in longer B2–B2 distances in the hexagonal rings (1.822(5) Å and 1.830(6) Å for $Nb_7Fe_3B_8$ and $Ta_7Fe_3B_8$, respectively) than the distances in $TMB_2$ (1.7962 Å and 1.7759 Å in $NbB_2$[67] and $TaB_2$[68], respectively), however, still shorter than the distances in the isolated hexagonal rings in $TM_7TM'_6B_8$ (1.868(4) Å, 1.870(6) Å and 1.84(4)–1.87(7) Å in $Nb_7Ru_6B_8$, $Ta_7Ru_6B_8$ and $Nb_7Rh_6B_8$).

The iron atoms in the crystal structure of $TM_7Fe_3B_8$ form a planar Kagome net at z = 1/2. With the relatively large Fe–Fe distances within the Kagome plane (4.17 Å for Nb and 4.14 Å for Ta compound) and the shorter distances of $d$(Fe–Fe) = 3.29 Å along [001] in both compounds (Fig. 4b), this atomic pattern plays important role in the magnetic behavior of these materials (cf. below).

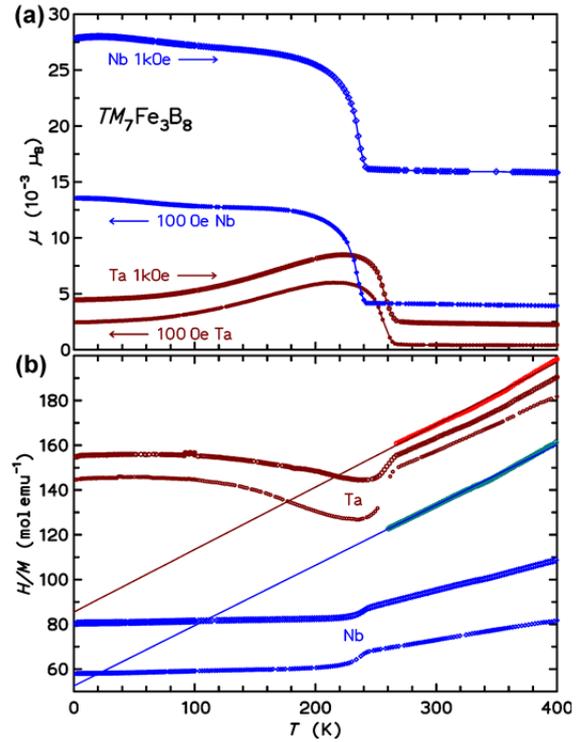

**Figure 6.** (a) Magnetic moment $\mu(T)$ measured in two different fields. The arrows indicate the direction of the temperature sweep. (b) Inverse susceptibility $H/M$ for $\mu_0H$ = 3.5 T (lower curves for $Nb_7Fe_3B_8$ and $Ta_7Fe_3B_8$, respectively) and 7.0 T (middle curves). The data sets above 250 K are the inverse intrinsic susceptibilities $1/\chi_{int}(T)$ (see text). The lines are Curie-Weiss fits to the latter data (see text).

**Physical Properties**

**Magnetic Properties:** The analysis of the magnetic properties of both $TM_7Fe_3B_8$ compounds is hampered by the presence of ferromagnetic impurities with high Curie temperatures. Fig. 6a shows the magnetic moment $\mu$ per formula unit (in Bohr magnetons $\mu_B$) for two low fields. Magnetic ordering transitions are clearly visible for both compounds, at $T_N$ = 240 K for $Nb_7Fe_3B_8$ and at $T_N$ = 265 K for $Ta_7Fe_3B_8$. Below these transitions, tiny remnant magnetization ≤ 0.012 $\mu_B$ for the Nb and 0.006 $\mu_B$ for the Ta compound are also observed. The bulk character of the magnetic transitions is confirmed by sizable anomalies in the specific heat (see below). Further transitions at low temperature are not observed.

Above the ordering transitions, the high-field paramagnetic susceptibilities $\chi(T) = M(T)/H$ were analyzed. First, the intrinsic susceptibility $\chi_{int}(T)$ was extrapolated by the Honda-Owen method from data taken in 3.5 T and 7 T field. Above 250 K, the $\chi_{int}(T)$ data are well described by the Curie-Weiss law (see the plot of $H/M$ in Fig. 6b), $\chi_{int}(T) = C/(T-\theta)$. The effective paramagnetic moment $\mu_{eff}$ calculated from $C$ and the Weiss parameter $\theta$ are 5.45 $\mu_B$ and −195 K for the Nb compound and 5.35 $\mu_B$ and −306 K for the Ta homologue.

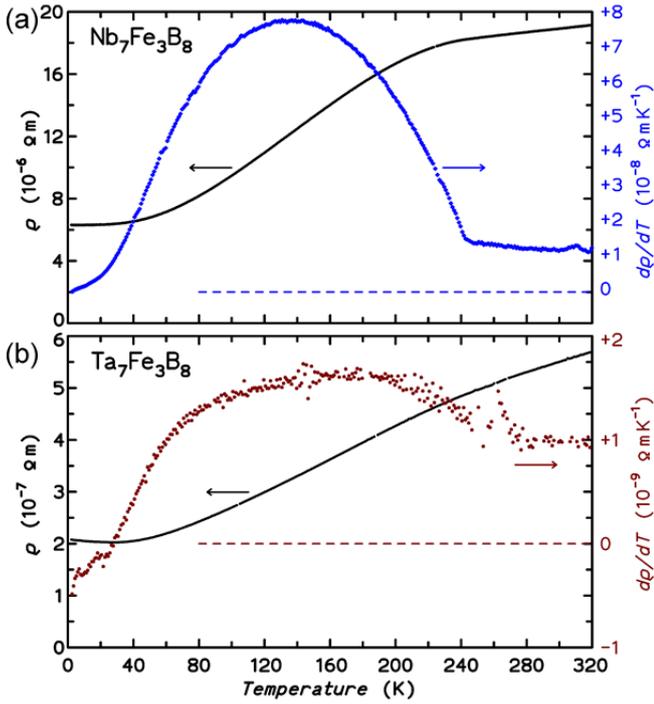

**Figure 7.** Electrical resistivity $\rho(T)$ and derivative $d\rho/dT$ for Nb$_7$Fe$_3$B$_8$ (a) and Ta$_7$Fe$_3$B$_8$ (b).

**Electrical Resistivity:** In Fig. 7, the electrical resistivities $\rho(T)$ and their temperature derivatives $d\rho/dT$ are shown. Both compounds display metallic conduction with $\rho_{300K} \approx 19$ μΩ m for the Nb and $\approx 0.55$ μΩ m for the Ta compound. While the former sample has an extraordinary large resistivity well above the Mott-Ioffe-Regel limit, the latter compound reaches only a value in the range typical for intermetallic compounds. For the Nb$_7$Fe$_3$B$_8$ a pronounced kink is visible at the magnetic transition. The kink at $T_N$ is weaker in the Ta compound. The derivatives $d\rho/dT$ indicate that for both compounds a contribution due to spin-disorder scattering of charge carriers is at work. Interestingly, the hump in $d\rho/dT$ of the Ta compound extends further to lower temperatures than that of Nb$_7$Fe$_3$B$_8$. The high residual resistivities (RRR value = 3.0 and 2.4 for Nb and Ta-containing compounds, respectively) suggest that both samples have considerable amounts of point defects.

**Specific Heat Capacity:** The isobaric specific heat $c_p(T)$ of the $TM_7$Fe$_3$B$_8$ compounds is shown in Fig. 8. The strongly bonded light boron atoms in these structures lead to high-frequency optical phonon modes. Therefore, the specific heat at room temperature is still well below the Dulong-Petit limit, i.e. $c_p \leq 3nN_Ak_B$ ($n$ = number of atoms in the formula unit, $N_A$ = Avogadro constant, $k_B$ = Boltzmann constant). There are clear second-order anomalies at the weak ferromagnetic ordering transitions. The anomaly for the Ta compound is smaller than that for the Nb homologue. Interestingly, $c_p(T)$ of Ta$_7$Fe$_3$B$_8$ is well above that of the Nb compound at temperatures below $\approx 140$ K, which may be expected from lower-lying phonon modes of the $TM$ species (Ta has almost twice the atomic mass of Nb).

The inset to Fig. 8 presents the low-$T$ specific heats in a $c_p/T$ vs. $T^2$ representation. For temperatures below 10 K the $c_p(T)$ may be analyzed following the ansatz $c_p(T) = aT^{-2} + \gamma T + \beta T^3$, where the first term captures the upturn towards the lowest temperatures (observed in the Ta compound only), $\gamma T$ is the contribution from conduction electrons, and $\beta T^3$ represents the Debye approximation of the lattice heat capacity. The Sommerfeld coefficients $\gamma$ are 32.9 and 38.0 mJ mol$^{-1}$ K$^{-1}$ for the Nb and Ta compound, respectively. The coefficients $\beta$ correspond to initial Debye temperatures $\theta_D$ of 688 K and 640 K, respectively, the lower $\theta_D$ of the latter compound being due to the large atomic mass of tantalum. The origin of the upturn at the lowest $T$ observed for Ta$_7$Fe$_3$B$_8$ is unclear. The application of magnetic fields $\mu_0 H$ of 3, 6, and 9 T leads to a progressive shift of entropy connected to this upturn to higher temperatures, however the involved entropy is very small compared to $N_A k_B$. This contribution is probably not due to hyperfine splitting of the nuclear multiplet of $^{181}$Ta ($I = 7/2$), especially since no corresponding effect is observed for Nb ($^{93}$Nb with $I = 9/2$).

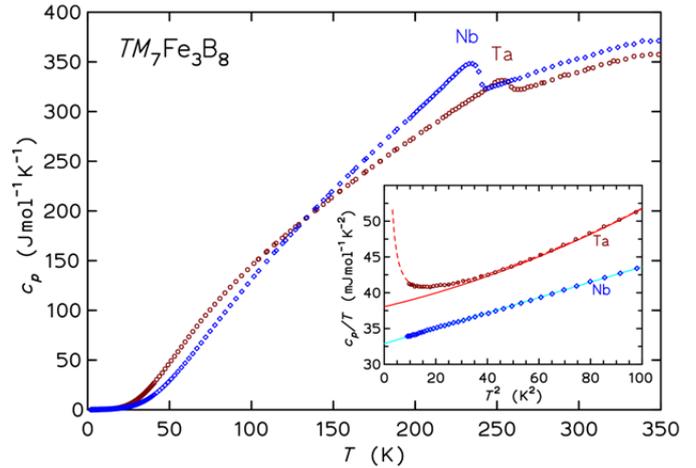

**Figure 8.** Specific heat capacity $c_p(T)$. The inset shows the low-$T$ range of the same data in $c_p/T$ vs. $T^2$ representation. The full lines are fits with electronic and lattice contributions, the dashed line in the case of Ta$_7$Fe$_3$B$_8$ also includes an $aT^{-2}$ contribution (see text).

**Mössbauer Spectroscopy:** Spectra of Nb$_7$Fe$_3$B$_8$ at 294 K and 4.3 K are shown in Fig. 9. The Mössbauer spectrum at 294 K shows a single line with small quadrupole splitting (Table 4). Magnetic hyperfine splitting is present at 4.3 K (Fig. 9) with a hyperfine field H$_{hf}$ less than one third of the value for pure α-Fe. The small line width (G/2) (Table 4) indicates the contribution of only one iron species in accordance with the crystal structure. The hyperfine splitting at 4.3 K confirms bulk nature of the magnetic order and suggests that all Fe nuclei experience similar hyperfine fields.

### Electronic structure

Both Nb$_7$Fe$_3$B$_8$ and Ta$_7$Fe$_3$B$_8$ are metallic. Their electronic density of states (DOS) features remarkably large contributions of $d$ states of both Nb/Ta and Fe at the Fermi level ($E_F$) (Fig. 10). According to the Stoner criterion band splitting and thus spontaneous ferromagnetic ordering are therefor very likely to occur. The Nb/Ta $4d$/$5d$ bands are very broad and span the energy range between −5 eV and 5 eV, where clear bonding (Nb and B) and antibonding (Nb only) combinations are formed below and above the Fermi level, respectively. The Fe $3d$ bands are about half as wide and weakly hybridize with the boron states. The electronic structures of the Nb and Ta compounds are very similar.

**Table 4.** Fitting parameters for Mössbauer spectra of Nb$_7$Fe$_3$B$_8$.

| $T$ (K) | $H_{hf}$(mm/s) | $eQV_{zz}/4$(mm/s) | CS(mm/s) | G/2(mm/s) |
|---|---|---|---|---|
| 294 | 0 | −0.020 | −0.033 | 0.158 |
| 4.3 | 3.611 | 0.038 | 0.058 | 0.126 |

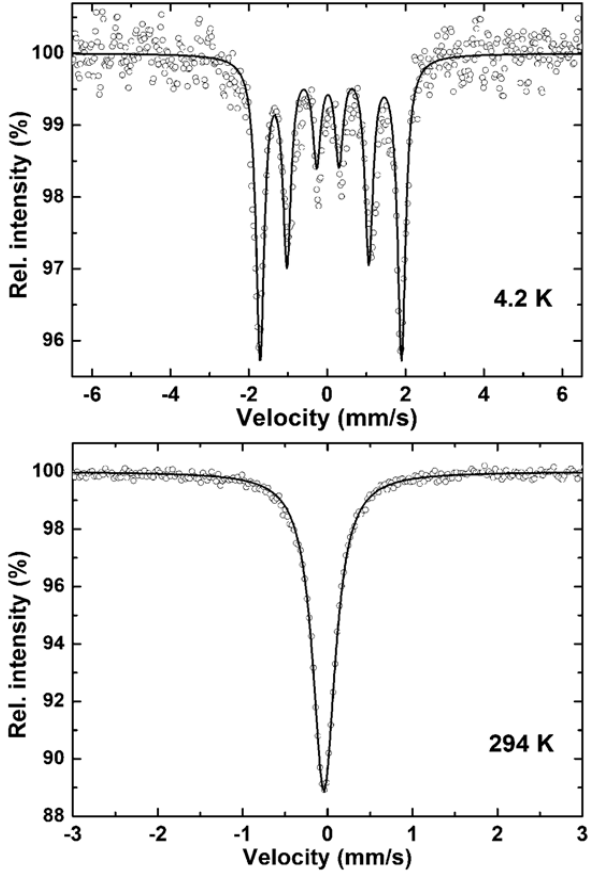

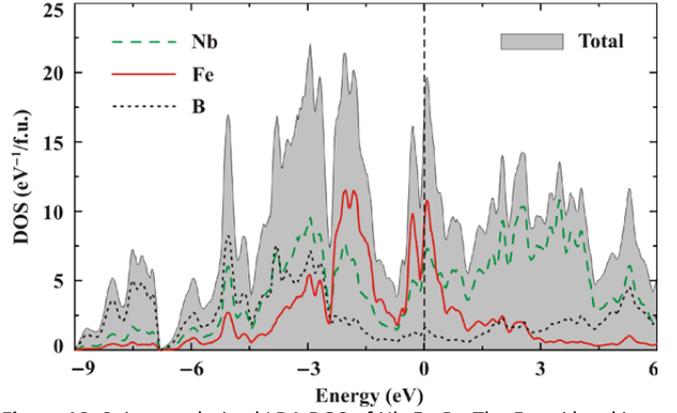

Figure 10. Spin-unpolarized LDA DOS of $Nb_7Fe_3B_8$. The Fermi level is at zero energy.

Figure 9. Mößbauer Spectra of $Nb_7Fe_3B_8$ at 4.3 K and 294 K.

The Fermi level is close to a local minimum of the DOS. The number of states at the Fermi level of about 16.5 eV$^{-1}$/f.u. and 13.5 eV$^{-1}$/f.u. in the spin-unpolarized calculation yields the Sommerfeld coefficients of $\gamma$ = 38 mJ mol$^{-1}$ K$^{-1}$ and 32 mJ/ mJ mol$^{-1}$ K$^{-1}$ for Nb$_7$Fe$_3$B$_8$ and Ta$_7$Fe$_3$B$_8$, respectively. These values are close to those found experimentally (32.9 and 38.0 mJ mol$^{-1}$ K$^{-1}$, respectively), but, surprisingly, reveal a different trend: $\gamma_{Nb}$ > $\gamma_{Ta}$ in DFT, whereas $\gamma_{Nb}$ < $\gamma_{Ta}$ experimentally. Spin fluctuations and magnetic order may affect the number of states at the Fermi level. However, this effect is difficult to assess computationally because the magnetic ground state of $TM_7Fe_3B_8$ is not known.

Spin-polarized calculations suggest that both Nb$_7$Fe$_3$B$_8$ and Ta$_7$Fe$_3$B$_8$ are magnetic. The magnetic moment on iron is about 1.72 $\mu_B$ regardless of the spin configuration. This value should not be confused with the paramagnetic effective moment of about 3.1 $\mu_B$/Fe, which is intrinsically higher than the ordered moment and reflects the full fluctuating spin moment, while 1.72 $\mu_B$ revealed by DFT is only its ordered part. On the other hand, this moment is much higher than the tiny remnant magnetization on the order of 0.01 $\mu_B$, which is seen below $T_N$.

In order to gain further insight into the nature of the magnetic order in $TM_7Fe_3B_8$, we analyzed nearest-neighbor exchange couplings by calculating total energies of several spin configurations. The following ordering patterns were considered: ferromagnetic order (I); ferromagnetic order in the $ab$ plane and antiferromagnetic order along $c$ (II); ferrimagnetic (up-up-down) order in the $ab$ plane and ferromagnetic order along $c$ (III). Note that we considered collinear spin configurations only. Therefore, a fully antiferromagnetic order in the $ab$ plane is not possible given the frustrated nature of the kagome spin lattice.

Our spin-polarized calculations revealed the lowest energy of configuration III that we further refer to as zero. The energies of the other two configurations are $E_I$ = +65.9 meV/f.u. and $E_{II}$ = +127.1 meV/f.u., respectively. This way, effective nearest-neighbor exchange couplings are $J_{ab}$ = 16.5 meV/Fe and $J_c$ = −21.2 meV/Fe, where the positive and negative signs stand for the antiferromagnetic and ferromagnetic couplings, respectively, and we do not divide energies by $S^2$ because in itinerant magnets it is not a good quantum number. We conclude that the couplings in the $ab$ plane are antiferromagnetic, while the coupling along $c$ is ferromagnetic. Therefore, despite large Fe–Fe distances (cf. above), $TM_7Fe_3B_8$ are magnetically frustrated, as no collinear spin configuration satisfies the antiferromagnetic couplings in the Kagome net. It is worth noting that effective exchange couplings of +190 K and −247 K are comparable in magnitude to the Curie-Weiss temperature of −195 K in Nb$_7$Fe$_3$B$_8$. However, both ferro- and antiferromagnetic interactions are observed.

**Chemical bonding analysis in real space**

A striking feature of the $TM_7Fe_3B_8$ crystal structure is the spatial separation of the $TM$ atoms from the Fe and B ones forming separated planar nets perpendicular to the [001] direction at $z$ = 0 and $z$ = ½ respectively. The reasons for such atomic arrangement were evaluated by the real space analysis of chemical bonding in $Nb_7Fe_3B_8$ employing electron localizability indicator ELI in its ELI-D representation[59] (Fig. 11). While the ELI-D distribution in the penultimate shell of boron atoms has a spherical shape as expected for a *p* element, the penultimate shells of niobium and iron atoms show strong inhomogeneity being the fingerprint of the participation of these electrons in the bonding interactions in the valence region.[60, 72] In the valence region around B1 atoms reveal five ELI-D maxima (attractors). Three of them are located on the Fe−B contacts visualizing the 2*c*(Fe−B) bonds. The basins of the remaining two attractors are located above and below the boron nucleus being in contact with the core basins of three Nb atoms beside ones of B1. This arrangement reflects the four-center interaction. ELI-D reveals similar distribution around B2. According to the local symmetry, two of the attractors in the plane at $z = ½$ visualize the 2*c*(B−B) bonds, the third one shows mainly 2*c*(Fe−B) interaction. The basins of the attractors above and below the plane are not present, being united with that of the Fe−B interaction indicating here a delocalization of a 2*c* bond toward a multicenter one. Thus, the plane of Fe and B atoms at $z = ½$ with its Kagome topology mentioned above is formed mainly by two-center interactions. Between this plane and the niobium atoms at $z = 0$ the multicenter bonding is observed. Assuming that the multicenter bonding is weaker than the two-center interactions, such bonding picture should yield a pronounced cleavage of the material perpendicular to the [001] direction.

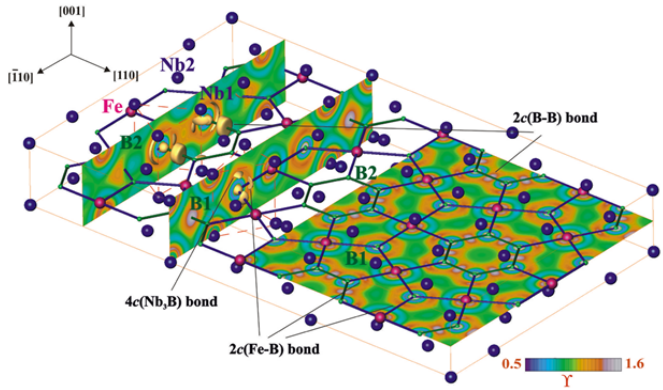

**Figure 11.** Electron localizability indicator in $Nb_7Fe_3B_8$: (left and middle) ELI-D ($\Upsilon$) distribution in the planes perpendicular to the Fe−B net revealing the structuring of the penultimate shells of Nb1, Nb2 and Fe atoms as well as location of the ELI-D attractors (shown by the yellow isosurface with $\Upsilon$ = 1.315) around the boron atoms within the [$BNb_6$] trigonal prisms (red dashed line); (right) ELI-D distribution in the plane at $z = ½$ with the maxima visualizing the two-center B−B and Fe−B bonds.

## Conclusions

In this study, two new ternary borides $TM_7Fe_3B_8$ ($TM$ = Nb, Ta) with Kagome-type iron sublattices were synthesized by arc-melting of the elements and subsequent annealing at 1500 °C. Their hexagonal primitive structure is an intergrowth of $AlB_2$-type and CsCl-type slabs, involving [$BTM_6$] trigonal prisms and [$FeTM_8$] tetragonal prisms. The condensation of trigonal prisms results in the formation of hexagonal columns along *c*-axis, hence, also forming isolated planar [$B_6$] rings in this structure.

Metallic character of $TM_7Fe_3B_8$ is confirmed by temperature dependence of the electrical resistivity as well as by the sizable linear term in the specific heat for both compounds.

Magnetic susceptibility measurements reveal predominantly antiferromagnetic order below $T_N$ = 240 K for $Nb_7Fe_3B_8$ and $T_N$ = 265 K for $Ta_7Fe_3B_8$. The sextet in the Mößbauer spectra of $Nb_7Fe_3B_8$, the sizable anomalies in the specific heat at $T_N$, and the kinks in the resistivity curves confirm the bulk character of the magnetic transitions for both compounds. These transitions are related to the presence of sizable magnetic moments localized on the Fe atoms within the planar Kagome-type iron sublattice. DFT calculations indicate metallic behaviour for both compounds and show that interactions in the *ab* plane are antiferromagnetic and thus subject to a strong geometrical frustration that should prevent Néel type magnetic ordering. On the other hand, strong interplane coupling (of any sign) can effectively suppress this frustration and trigger the formation of long-range-ordered states,[73] which is probably the case in $TM_7Fe_3B_8$. However, the presence of remnant magnetization unanticipated in a regular Kagome antiferromagnet indicates a more complex nature of the magnetic order.

Quantum-chemical analysis of the chemical bonding in $Nb_7Fe_3B_8$ within the electron localizability approach reveals five ELI-D maxima around B1, visualizing three in-layer Fe−B bonds and two interactions with core basins of three Nb atoms above and below, while only three maxima around B1, showing two B−B bonds and one Fe−B interaction. The analysis also indicates that the Kagome net of Fe and B is mainly formed by two-center interactions, whereas multicenter bonding between this net and Nb atoms is observed.

## Acknowledgements

We thank Mr. S. Hückmann and Dr. Yu. Prots for performing powder X-ray diffraction measurements, Ms. P. Scheppan, Ms. M. Eckert and Ms. S. Kostmann for metallographic analysis, and Mr. R. Koban for physical property measurements. We are grateful to Dr. G. Auffermann and Ms. U. Schmidt for ICP-MS analysis of the samples. We thank Dr. K. Hradil for preliminary low temperature powder XRD measurements at the X-ray center of the TU Wien. AT was supported by the Federal Ministry for Education and Research through the Sofja Kovalevskaya award of Alexander von Humboldt Foundation.